\documentclass[pra,aps,showpacs,superscriptaddress,twocolumn]{revtex4}
\usepackage{graphicx}
\usepackage{amsmath}
\usepackage{amsfonts}
\usepackage{amssymb}
\usepackage{epsfig}
\usepackage{color}

\begin{document}

\newcommand{\abs}[1]{\left\vert#1\right\vert}
\newcommand{\set}[1]{\left\{#1\right\}}
\newcommand{\bra}[1]{\left\langle#1\right\vert}
\newcommand{\ket}[1]{\left\vert#1\right\rangle}
\newcommand\braket[2]{\left.\left\langle#1\right|#2\right\rangle}
\def\I {{\rm 1} \hspace{-1.1mm} {\rm I} \hspace{0.5mm}}
\newcommand{\rosso}[1]{\color[rgb]{0.6,0,0} #1}

\title{Quantum Zeno and anti-Zeno effects on quantum and classical correlations}

\author{F. Francica}\email{francesco.francica@fis.unical.it}
\affiliation{Dipartimento di  Fisica, Universit\`a della Calabria,
87036 Arcavacata di Rende (CS) Italy} \affiliation{INFN - Gruppo
collegato di Cosenza}
\author{F. Plastina}\email{francesco.plastina@fis.unical.it}
\affiliation{Dipartimento di  Fisica, Universit\`a della Calabria,
87036 Arcavacata di Rende (CS) Italy} \affiliation{INFN - Gruppo
collegato di Cosenza}
\author{S. Maniscalco}\email{sabrina.maniscalco@utu.fi}\homepage[]{www.openq.fi}
\affiliation{Turku Centre for Quantum Physics, Department of Physics and Astronomy, University of
Turku, FI-20014 Turun yliopisto, Finland}

\date{\today}

\begin{abstract}
In this paper we study the possibility of modifying the dynamics
of both quantum correlations, such as entanglement and discord,
and classical correlations of an open bi-partite system by means
of the quantum Zeno effect. We consider two qubits coupled to a
common boson reservoir at zero temperature. This model describes,
for example, two atoms interacting with a quantized mode of a
lossy cavity. We show that,  when the frequencies of the two atoms
are symmetrically detuned from that of the cavity mode,
oscillations between Zeno and anti-Zeno regimes occur. We also
calculate analytically the time evolution of both classical
correlations and quantum discord, and we compare the Zeno dynamics
of entanglement with the Zeno dynamics of classical correlations
and discord.

\end{abstract}

\bigskip

\maketitle
\section{Introduction}
Composite quantum systems may possess correlations of a different
nature with respect to classical ones. For the latter,
correlations are generally measured by the mutual information,
whose extension to the quantum realm, however, leads to two
different quantities: the quantum mutual information $\mathcal{I}$
and the classical correlations $\mathcal{C}_c$
\cite{OlliZurek01,HendVed01}. The quantum discord is the
difference between these two quantities, i.e.,
$\mathcal{D}=\mathcal{I}-\mathcal{C}_c$. This quantity is zero
only for classically correlated states, i.e., in the case of
bipartite systems, for states of the form $\rho = \sum_{k,l}
p_{k,l} \vert k,l \rangle \langle k,l \vert$, with $p_{k,l} \neq
p_k p_l$, $\vert k \rangle$ and $\vert l \rangle$ being orthogonal
states of the two subsystems. In this sense the quantum discord
measures quantum correlations.

For mixed quantum states, the discord does not coincide with
entanglement. Indeed, there exist separable states having non-zero
discord \cite{Luo}. Recently, the properties of quantum and
classical correlations have received a huge deal of attention in
both applicative and fundamental research \cite{Luo, Opp, Grois,
Moelmer, Modi, Zurek03, Horodec, Rodriguez, Datta, Piani,
Maziero0, Shabani, Datta2, Piani2, Werlang, Fanchini, Maziero,
Ferraro, Lanyon}. In particular, it has been demonstrated that
states having nonzero discord but zero entanglement can be used in
certain models of quantum computation to achieve significant
speedup with respect to the classical algorithms \cite{Datta,
Datta2, Lanyon}. Moreover, it has been shown that, under certain
conditions, quantum discord is totally unaffected by
non-dissipative noise for long time intervals \cite{LauraPRL,Laurapreprint}.

In this article we study the effect of nonselective projective
measurements on the dynamics of a simple bipartite quantum system.
Our aim is to investigate how both quantum and classical
correlations are affected by the measurements. We focus, in
particular, on a system of two noninteracting qubits immersed in a
common structured reservoir, such as the principal mode of a
high-Q cavity. It is known that, in this case, certain types of
measurements performed on either the collective state of the
system or the state of the cavity, may inhibit the loss of quantum
entanglement initially present in the two qubits states
\cite{man08,kurrp}. The resulting measurement-induced protection
of entanglement was studied in the case in which the two qubits
are resonantly coupled with the cavity mode.

We now extend the results we have presented in a previous Letter
\cite{man08} in two directions. First of all we consider different
cases of off-resonant interaction between the qubits and the
cavity mode and we show that, under certain conditions, the same
measurements that allowed to significantly suppress entanglement
loss are now responsible for a much faster disentanglement, as a
result of the anti-Zeno effect \cite{kof00,facch}. Secondly, we
study how measurements affect not only entanglement but also other
types of correlations between the qubits, namely classical
correlations and quantum discord. We show that, contrarily to what
has been found in previously studied systems \cite{Maziero0,
Werlang, Maziero}, the dynamics of classical and quantum
correlations, in the system here considered, is qualitatively very
similar to the dynamics of entanglement. Therefore, projective
measurements have a very similar effect on these three types of
correlations. In particular, we show that there exist working
conditions for which a series of oscillations occurs between the
Zeno and anti-Zeno regimes, both for the quantum and the classical
correlations, and that the amount of correlations can depend on
the relative phase of the initial state. Oscillations between Zeno and anti-Zeno regimes have also been predicted in the quantum Brownian motion model for certain types of structured environments \cite{Janika2010}.

The plan of the paper is the following. In Sec. II we describe the
physical model employed in Sec. III to derive the analytic
expression of classical correlation and of quantum discord. In
Sec. IV, we obtain analytically the time evolution of the system
in presence of projective non-selective measurements on the
qubits, which are then used in Sec. V, where we present the
quantum Zeno and anti-Zeno effects on concurrence, discord and
classical correlations. Finally, Sec. VI provides a summary of the
results together with some concluding remarks.

\section{The System}

Let us consider an
open quantum system consisting of two  qubits coupled to a common
zero-temperature bosonic reservoir. The Hamiltonian of the total system is
\begin{equation}\label{eq:H}
    H=H_S+H_R+H_{\rm int},
\end{equation}
where $H_R$ is the Hamiltonian of the reservoir and $H_S$ is the
Hamiltonian of the two qubits which are coupled to the common
reservoir via the interaction $H_{\rm int}$.

The Hamiltonian for the total system, in the dipole and
the rotating-wave approximations, and in units of $\hbar$, reads
\begin{eqnarray}
H_S &=& \omega_1 \sigma^{(1)}_+  \sigma^{(1)}_- + \omega_2
\sigma^{(2)}_+\sigma^{(2)}_-, \\
H_R &=& \sum_k \omega_k b_k^{\dag} b_k, \\
H_{\rm int} &=&  \left( \alpha_1 \sigma^{(1)}_+   +
\alpha_2\sigma^{(2)}_+ \right) \sum_k g_k b_k  + {\rm h.c.},
\label{eq:Hint}
\end{eqnarray}
where $b^\dag_k$, $b_k$ are the creation and annihilation
operators of quanta of the reservoir,  $ \sigma^{(j)}_{\pm}$ and
$\omega_{j}$ are the inversion operator and transition frequency
of the $j$-th qubit  ($j=1,2$), $\omega_k$ are the frequencies of
the reservoir $k$-mode, and
 $\alpha_j g_k$ describe the coupling strength between the
$j$-th qubit and the $k$-mode of reservoir.


The $\alpha_j$ are dimensionless real coupling constants
measuring the interaction strength of each single qubit with the
reservoir. We assume that the ratio between these two constants can be
varied independently. In the case of two atoms inside a cavity,
e.g., this can be achieved by changing the relative position of the
atoms with respect to the the cavity field standing wave. We denote with
$\alpha_T=(\alpha_1^2+\alpha_2^2)^{1/2}$ the collective coupling
constant and with $r_j=\alpha_j/\alpha_T$ the relative interaction
strength.

We restrict ourselves to the case in which only one excitation is
present in the system and the reservoir is in the vacuum.
Initially the two-qubit system is assumed to be disentangled from
its reservoir and the initial state for the whole system is
written as
\begin{equation}
\ket{\Psi(0)} = \Bigl [ c_{01} \ket{1}_1\ket{0}_2 + c_{02}
\ket{0}_1\ket{1}_2\Bigr] \bigotimes_k
\ket{0_k}_R,\label{initialstate}
\end{equation}
where $c_{01} = \sqrt{(1-s)/2}$ and $c_{02}=\sqrt{(1+s)/2} e^{i
\phi}$ are complex numbers defining the initial state for the
qubits system, $-1 \le s \le 1$,
 $\ket{0}_j$ and $\ket{1}_j$ $(j=1,2)$ are the ground and excited state of the $j$-th qubit,
respectively, and $\ket{0_k}_R$
is the state of the reservoir with zero excitations in the $k$-mode.

As a consequence of the time evolution generated by the Hamiltonian
(\ref{eq:Hint}), the excitation can be shared by the qubits and
the reservoir, so that
\begin{eqnarray}
\vert \Psi (t) \rangle &=&  c_1 (t) \vert 1 \rangle_1 \vert 0
\rangle_2 \vert 0 \rangle_R + c_2(t)  \vert 0 \rangle_1 \vert 1
\rangle_2 \vert 0 \rangle_R+ \nonumber \\ &&+ \sum_k c_k (t) \vert
0 \rangle_1 \vert 0 \rangle_2 \vert 1_k \rangle_R, \label{eq:psi}
\end{eqnarray}
$\vert 1_k \rangle_R$ being the state of the reservoir with only
one excitation in the $k$-th mode.

In the standard basis, the reduced density matrix for the qubits,
obtained from the density operator $\vert \Psi(t) \rangle \langle
\Psi(t) \vert$ after tracing over the reservoir degrees of
freedom, takes the form
\begin{equation}
\label{eq:rhos} \rho(t) = \left(
  \begin{array}{cccc}
    1-|c_1(t)|^2-|c_2(t)|^2& 0 & 0 & 0 \\
    0&|c_2(t)|^2&c_1^*(t)c_2(t)& 0\\
    0&c_1(t)c_2^*(t)&|c_1(t)|^2& 0\\
    0&0&0&0\\
  \end{array}
\right).
\end{equation}
The two-qubit dynamics is therefore completely characterized by
the amplitudes $c_{1,2}(t)$. For certain specific structures of
the reservoir, one can obtain the exact analytical expressions of
$c_{1,2}(t)$ by the Laplace transform method. In this paper we
consider a structured reservoir describing the electromagnetic
field inside a lossy cavity. This case can be modelled by a
Lorentzian broadening of the fundamental  cavity mode. The
non-Markovian analytical expression for the amplitudes
$c_{1,2}(t)$, in the off-resonant regime, were presented in Ref.
\cite{fran09}. We make explicit use of these results here. As a specific example of application to a real system, this model
has been shown to adequately describe the dynamics of ions trapped
in an electromagnetic resonator \cite{Harkonen}.

\section{Classical  and quantum correlations}
In this section we recall the analytic expression for the
entanglement dynamics, as measured by concurrence, and present the
analytic formula for both the classical correlations and the
discord. The entanglement dynamics for a generic initial two-qubit
state containing one excitation coupled to a common structured
reservoir was investigated in \cite{man08,fran09}. We choose the
concurrence $C_E(t)$ \cite{wootte}, ranging from $0$ for separable
states to $1$ for maximally entangled states, to quantify the
amount of entanglement encoded into the two-qubit system. This
quantity can be obtained from the reduced density matrix of Eq.
(\ref{eq:rhos}) and takes the simple form
\begin{equation}
\label{Ct}
   C_E\left(t\right)=2 \abs{c_1(t)} \,\abs{c_2(t)}.
\end{equation}

In Ref. \cite{man08} we have shown how, in the resonant regime,
repeated nonselective measurements on the collective state of the
qubits system induce a quantum Zeno effect \cite{Misra77} on the
entanglement and we have proven that, in this way, one can protect
entanglement. In the following section we investigate whether the
same occurs when the two qubits are off-resonant with the cavity
mode.

We start by focussing on the classical correlations present in the
system. In a bipartite quantum state of two qubits, classical
correlations, are defined as \cite{OlliZurek01,HendVed01}
\begin{eqnarray}
\label{eq:clascorr}
   \mathcal{C}_c(\rho) &=& \sup_{\{ \Pi_k^{(2)} \}} \left[ S(\rho_1)-S(\rho | \{\Pi_k^{(2)}
  \})\right],
\end{eqnarray}
where the maximum is taken over all projective measurements
performed locally on qubit $2$, described by a set of orthogonal
projectors $\{\Pi_k^{(2)} \}$ corresponding to the outcomes $k$.
In Eq. (\ref{eq:clascorr}), $S(\rho)$ is the von Neumann entropy,
$\rho_1$ the reduced density operator of qubit $1$, and $S(\rho |
\{\Pi_k^{(2)} \})$ the conditional entropy defined as $S(\rho |
\{\Pi_k^{(2)} \})=\sum_k p_k S(\rho_k)$, with $\rho_k=[ (I^{(1)}
\otimes \Pi_k^{(2)})\rho (I^{(1)} \otimes \Pi_k^{(2)})]/p_k$  the
conditional density operator of qubit 1 after qubit $2$ is measured
and the outcome $k$ is obtained, with probability $p_k=Tr
[(I^{(1)} \otimes \Pi_k^{(2)}) \rho (I^{(1)} \otimes
\Pi_k^{(2)})]$.

For the system considered in this paper, the optimization problem
in the definition of the classical correlations can be solved
exactly and a simple analytical expression for this quantity can
be derived. Indeed, by calculating the action of the one-qubit
projectors
\begin{eqnarray}\label{projQubit2}
  \Pi_k^{(2)} &=& I\otimes \ket{k}\bra{k}, \qquad \mathrm{with} \quad
    k=a,b
\end{eqnarray}
and
\begin{eqnarray*}
  \ket{a} &=& \cos \theta \ket{\uparrow}+e^{i \phi} \sin \theta \ket{\downarrow}, \\
  \ket{b} &=& \sin \theta \ket{\uparrow}-e^{i \phi} \cos \theta \ket{\downarrow},
\end{eqnarray*}
on the general two-qubit state given by Eq. (\ref{eq:rhos}), and
using Eq. (\ref{eq:clascorr}), it is straightforward to prove that
the classical correlations do not explicitly depend on $\phi$ and
are maximized for $\theta=n \, \pi/2$ with $n\in \mathbb{Z}$. The
analytic expression for $\mathcal{C}_c$ is given by
\begin{eqnarray}\label{classCorr}
  \nonumber \mathcal{C}_c(\rho) &=& \left(1-|c_1(t)|^2-|c_2(t)|^2 \right)\log_2 \left(1-|c_1(t)|^2-|c_2(t)|^2
  \right)\\
  \nonumber &-& \sum_{j=1,2}\left(1-|c_j(t)|^2 \right)\log_2 \left(1-|c_j(t)|^2 \right).\\
\end{eqnarray}

The dynamics of the  quantum discord is then easily calculated as
$\mathcal{D}(t)=\mathcal{I}(t)-\mathcal{C}_c(t)$, with
\begin{equation}\label{mutual}
{\cal I}(\rho)=S(\rho_1)+S(\rho_2)-S(\rho).
\end{equation}
The analytic expression of the discord reads
\begin{eqnarray}\label{QuantDisc}
  \nonumber \mathcal{D}(\rho) &=& |c_1(t)|^2 \, \log_2 \left(1+\frac{\, |c_2(t)|^2}{\, |c_1(t)|^2}
  \right) \\
  && + \,  |c_2(t)|^2 \, \log_2 \left(1+\frac{\, |c_1(t)|^2}{\, |c_2(t)|^2} \right)
\end{eqnarray}
We note that, if times $\bar{t}$ such that
$|c_1(\bar{t})|=|c_2(\bar{t})|$ exist, then
$\mathcal{D}(\bar{t})=C_E(\bar{t})$, i.e., the quantum
correlations as measured by the discord coincide with entanglement
as measured by concurrence, although the state is not necessarily
pure.

In fact, one can show in general that, for any two-qubit density
matrix of the form $\rho = (1-\alpha) |00\rangle \langle 00| +
\alpha |\psi_{me}\rangle \langle \psi_{me}|$, where
$|\psi_{me}\rangle$ is any maximally entangled state orthogonal to
$|00\rangle$, and $\alpha \in [0,1]$, the concurrence is equal to
the discord: $C_E = {\cal D} = \alpha$. The case discussed here
(with $|c_1|=|c_2|$) gives precisely a density matrix of the
previous kind, with $\alpha= 2 |c_1|^2$.

We can conclude, therefore, that, in these cases, the system does
not contain quantum correlations other than entanglement. We also
note that, for $|c_1(\bar{t})|=|c_2(\bar{t})|$, the discord
recently defined by Modi et al. in terms of the distance to the
closest classically correlated state \cite{Modi}, coincides with
the one used in this paper.

In the next section we will study how appropriately designed
nonselective projective measurements modify the dynamics of
quantum and classical correlations. Contrarily to what happens in
the resonant case \cite{man08}, we will see that such type of
measurements can enhance rather than protect the decay of quantum
correlations, and that even classical correlations are affected in
a similar way.

As peculiar specific instances in the rich parameter space of our
model, we will give special attention to two possible
configurations; namely, those corresponding to detunings
$\delta_1=\pm \delta_2$ ($\delta_{i}$ being the detuning of the
$i$-th atom from the cavity mode), with equal couplings between
atoms and reservoir, i.e., $r_1=r_2$. This choice stems from the
analysis of the system dynamics in absence of measurements
performed in Ref. \cite{fran09}. We have seen there that these two
cases give rise to interesting dynamical behavior and therefore
deserve special attention.

From the analytic expressions of the coefficients $c_1(t)$ and
$c_2(t)$ derived in Ref. \cite{fran09} one can prove that for
$\delta_1=\pm \delta_2$ and $r_1=r_2$,  $|c_1(t)|=|c_2(t)|$ at all
times $t$. To the best of our knowledge this is the first example
of open system dynamics for which during the whole time evolution
the quantum correlations exactly coincide with entanglement. We
recall that this is always the case for pure states (with the
entanglement measured by the entropy), but for mixed states as
those of our open system, this is far from being trivial.

\section{The effect of projective measurements}

We recall that, in order to observe the quantum Zeno effect on the
entanglement, the series of nonselective measurements on the
collective atomic system, performed at time intervals $T$, must have
the two following properties: i) one of the possible measurement
outcomes is the projection onto the collective ground state $\vert
\psi_0 \rangle = \vert 0 \rangle_1 \vert 0 \rangle_2$, and ii) the
measurement cannot distinguish between the excited-states
$\ket{\psi_1}=\ket{1}_1 \ket{0}_2$ and
$\ket{\psi_2}=\ket{0}_1\ket{1}_2$.

Such measurements are described by the following two projectors:
\begin{eqnarray}
  \Pi_0 &=& \ket{\psi_0}\bra{\psi_0}\otimes
    I_R, \\
  \Pi_1 &=& \left( \ket{\psi_1}\bra{\psi_1}+\ket{\psi_2}\bra{\psi_2}\right)\otimes
    I_R, \label{Pi1}
\end{eqnarray}
with $I_R$ the reservoir identity matrix. Projective measurements as those described by the operator $\Pi_0$
can be experimentally implemented in both cavity QED
\cite{QEDsetup} and in superconducting circuits with on-chip
qubits and resonator \cite{Sill07,Maj07}.

The state of the total system, formed by the two-qubits and the
electromagnetic field inside the cavity, after a series of $N$
instantaneous ideal measurements performed at time intervals $T$
is given by
\begin{eqnarray}\label{psi12N}
\vert \Psi^{(N)} (t) \rangle &=&  \Bigl [c_1^{(N)} (T) \ket{\psi_1}
+ c_2^{(N)}(T) \ket{\psi_2}\Bigr]\bigotimes_k \vert 0_k \rangle_R
\nonumber \\
&&+\sum_k b_k^{(N)}(T)\ket{\psi_0}\vert 1_k \rangle_R,
\end{eqnarray}
$c_{1,2}^{(N)}(T)$ and $b_k^{(N)}(T)$ are the survival amplitudes
at time $t=NT$ related to the presence of the excitation in qubit
1, qubit 2, and the cavity field, respectively.

We note that the reduced density matrix $\rho^{(N)}(t)= Tr
\left\{\ket{\Psi^{(N)} (t)}\bra{\Psi^{(N)} (t)}\right\}_R$,
describing the two-qubit system after $N$ measurements, has the
same structure of Eq. (\ref{eq:rhos}), provided one changes
$c_{1,2}(t)$ with  $c_{1,2}^{(N)}(T)$.

Our aim is to derive simple and physically transparent equations
for  $c_{1,2}^{(N)}(T)$ able to explain the occurrence of quantum
Zeno or anti-Zeno effects for classical and quantum correlations.
To achieve this goal we analyze first the coarse grained evolution
with $t=NT$, and then derive, under certain approximations, the
time evolution of the system between successive measurements.

\subsection{Coarse grained dynamics}

Let us introduce the matrix
$\mathbf{E}$ describing the uninterrupted evolution between two
consecutive measurements in the two-qubit subspace spanned
by $\ket{\psi_1}$ and $\ket{\psi_2}$. In the following we focus on the case of frequently observed
dynamics and we assume that the interval $T$ between two
successive measurements is short. The survival amplitudes
$c_{1,2}^{(N)}(T)$ in presence of measurements can be written as
\begin{equation}\label{c12N}
\begin{array}{cccc}
    \left(
       \begin{array}{c}
         c_1^{(N)}(T) \\
         c_2^{(N)}(T) \\
       \end{array}
       \right)
   & = & \mathbf{E}^{(N)}
   & \left(
       \begin{array}{c}
         c_1(0) \\
         c_2(0) \\
       \end{array}
     \right)
\end{array}
\end{equation}
where
\begin{equation}\label{matrixEN}
\mathbf{E}^{(N)}=\left(
    \begin{array}{cc}
      \mathcal{E}_{11}^{(N)}(T) & \mathcal{E}_{12}^{(N)}(T) \\
      \mathcal{E}_{21}^{(N)}(T) & \mathcal{E}_{22}^{(N)}(T) \\
    \end{array}
  \right),
\end{equation}
$\mathbf{E}^{(N)}$ being the evolution matrix in presence of $N$
measurements. Note that, in general, $\mathbf{E}^{(N)}$ is not
equal to the $N$-th power of the evolution matrix $E$ between two
successive measurements; therefore, the explicit expressions of
the matrix elements $\mathcal{E}_{ij}(T)$, obtained by the Laplace
transform method \cite{fran09}, take a complicated form.

Simple forms for the survival amplitudes in presence of
measurements can be found when $\omega_1=\omega_2$, since in this
case only the superradiant state evolves \cite{fran09}. In this
case, it is much more useful to express the matrix
$\mathbf{E}^{(N)}$ in the superradiant-subradiant basis, so that
it has only one non-zero entry, given by the survival amplitude of
the superradiant state, which takes the form
\begin{equation}\label{EvoPM}
{\cal E}(T) = e^{- (\lambda -\, i \delta)\, T /2} \left[ \cosh \left
( \frac{\Omega T}{2}\right) + \frac{\lambda - i \delta}{\Omega} \,
\sinh \left( \frac{\Omega T}{2} \right) \right],
\end{equation}
where $\delta_1=\delta_2\equiv\delta$,
$\Omega=\sqrt{\lambda^2-\Omega_R^2-i 2\delta \lambda}$, and where
we indicate with $\lambda$ the width of the Lorentzian spectral
distribution describing the field inside the cavity. Here,
$\Omega_R=\sqrt{4 W^2 \alpha_T^2+\delta^2}$ is the generalized
Rabi frequency, while $\mathcal{R}=W \alpha_T$ is the vacuum Rabi
frequency.

For the general case,  one can prove that, if $\lambda T\ll 1$,
i.e., in the limit of frequent measurements, the off-diagonal
elements of the evolution matrix $\mathcal{E}_{ji}(T)$, with
$j\neq i$, are small and decrease quickly as $T$ decreases and
$\delta_{1,2}$ increases. Hence, the dynamics for long time
intervals are well described by the following expressions
\begin{eqnarray}
&& \nonumber \label{EvoNsum1} \mathcal{E}_{jj}^{(N)}(T) \approx \mathcal{E}_{jj}^N(T)\Big[ 1+\theta\left[ N-1 \right] \frac{\mathcal{E}_{ji}(T)\mathcal{E}_{ij}(T)}{\mathcal{E}_{jj}(T)^2}  \\
 && \times \sum_{k=0}^{N-2}(N-1-k)\left(\frac{\mathcal{E}_{ii}(T)}{\mathcal{E}_{jj}(T)}\right)^k \Big],
\end{eqnarray}
\begin{eqnarray}
&& \nonumber  \label{EvoNsum2}  \mathcal{E}_{ji}^{(N)}(T) \approx \mathcal{E}_{jj}^N(T)\Big[ \frac{\mathcal{E}_{ji}(T)}{\mathcal{E}_{jj}(T)}\sum_{k=0}^{N-1}\left(\frac{\mathcal{E}_{ii}(T)}{\mathcal{E}_{jj}(T)} \right)^k+\theta\left[ N-2 \right]\\
  &&\times \frac{\mathcal{E}_{ji}^2(T) \mathcal{E}_{ij}(T)}{\mathcal{E}_{jj}(T)^3} \sum_{k=0}^{N-3}(k+1)(N-k)\left(\frac{\mathcal{E}_{ii}(T)}{\mathcal{E}_{jj}(T)}
  \right)^k \Big],
\end{eqnarray}
where $\theta[x]$ is the Heaviside step function.

Inserting Eqs.(\ref{EvoNsum1})-(\ref{EvoNsum2}) into Eqs.
(\ref{c12N})-(\ref{matrixEN}) we obtain the dynamics of the
two-qubits reduced density matrix  $\rho^{(N)}(t)$ at time $t=NT$
in presence of measurements, which is found to retain the same
structure as in Eq. (\ref{eq:rhos}). As a consequence, the
concurrence $ C_E^{(N)}(t)$, the classical correlations $
\mathcal{C}_c^{(N)}(t)$ and the discord $ \mathcal{D}^{(N)}(t)$,
in presence of measurements, are simply obtained by replacing the
amplitudes $c_{1,2}(t)$ with $c_{1,2}^{(N)}(T)$  in Eq.
(\ref{Ct}), Eq. (\ref{classCorr}) and Eq. (\ref{QuantDisc}),
respectively.

The dynamics of quantum and classical correlations in presence of
measurements depends on $T$, on the initial state, on the
detunings $\delta_{1,2}$,  on the relative coupling between the
qubits and the cavity field, and on the quality factor of the
cavity, i.e. on $\lambda$. To study in detail how such factors
influence the dynamics we need to evaluate analytic expressions
for $\mathcal{E}_{ij}(T)$. We will face this task in next
subsection.

\subsection{Evolution between two successive measurements}

In the limit of frequent measurements, Eqs.
(\ref{EvoNsum1})-(\ref{EvoNsum2}) imply that the evolution in
presence of $N$ measurements can be obtained from the matrix
elements $\mathcal{E}_{ij}(T)$. Therefore we need only to
calculate, in this limit, the time evolution of the amplitudes
$c_j(t)$, in the interval $0 \le t \le T$ between two successive
measurements. To this aim we employ perturbation theory, valid
when the evolution time $T$ is sufficiently short, such that
$c_j(T)\approx c_{j}(0)\equiv c_{j0}$ with $j=1,2$ \cite{kof00}.

The integro-differential equations for the probability amplitudes,
obtained from the Schr\"odinger equation \cite{fran09}, are
\begin{eqnarray}\label{eq:Dotcjf}
  \nonumber \dot{c}_j(t) &=& -\alpha_j^2 \, \int_0^t \! dt^\prime   f(t^\prime) e^{i\, \delta_j t^\prime} \, c_j(t-t^\prime) \\
   & -& \alpha_j \alpha_i \, e^{i \, (\delta_{j}-\delta_{i}) t} \,\int_0^t \! dt^\prime   f(t^\prime) e^{i\, \delta_i t^\prime} \, c_i(t-t^\prime),
\end{eqnarray}
with $j\neq i$. To first order, one gets
\begin{eqnarray}\label{eq:DotcjfAppx}
  \nonumber c_j(T) &=&c_{j0} -\alpha_j^2 \, \int_0^T \! dt \int_0^t \! dt^\prime   f(t^\prime) e^{i\, \delta_j t^\prime} \, c_{j0} \\
   & -& \alpha_j \alpha_i \,\int_0^T \! dt e^{i \, (\delta_{j}-\delta_{i}) t} \,\int_0^t \! dt^\prime   f(t^\prime) e^{i\, \delta_i t^\prime} \,
   c_{i0},
\end{eqnarray}
where $\delta_j=\omega_j-\omega_c$, $\omega_c$ is the fundamental
frequency of the cavity, and $f(t)$ is the correlation
function.

Recalling that the correlation function is the Fourier
transform of the reservoir spectral density $J(\omega)$,
\begin{equation}\label{corfunc}
    f(t)=\int \! d\omega J(\omega)
e^{i(\omega_c-\omega)t},
\end{equation}
we can recast the amplitudes of Eq. (\ref{eq:DotcjfAppx}) in the form
\begin{eqnarray}\label{eq:cjOverlap}
  \nonumber c_j(T) &=&c_{j0} -\alpha_j^2 \, \int \! d\omega J(\omega) F_{jj}(\omega,T) \, c_{j0} \\
   && -\alpha_j \alpha_i \int \! d\omega J(\omega) F_{ji}(\omega,T) \,
   c_{i0}.
\end{eqnarray}
This equation shows how the amplitudes $ c_j(T)$ depend on two
form factors, $F_{jj}(\omega, T)$ and $F_{ji}(\omega, T)$, defined
as
\begin{eqnarray*}
F_{jj}(\omega, T)&=& \int_0^T \! d t \int_0^\infty \! d t^\prime \, \theta[t-t^\prime] \, e^{i \omega_j t^\prime} e^{-i \omega t^\prime} \\
&=&\frac{1-e^{i (\omega_j-\omega)T}+i
(\omega_j-\omega)T}{(\omega_j-\omega)^2},
\end{eqnarray*}
and
\begin{eqnarray*}
F_{ji}(\omega, T)&=& \int_0^T \! d t \, e^{i (\delta_j-\delta_i)
t} \int_0^\infty \! d t^\prime \,
\theta[t-t^\prime] \, e^{i \omega_i t^\prime} e^{-i \omega t^\prime} \\
&=&\frac{1-e^{i(\omega_j-\omega)T}}{(\omega_j-\omega)(\omega_i-\omega)}
-\frac{1-e^{i
(\omega_j-\omega_i)T}}{(\omega_i-\omega)(\omega_j-\omega_i)},
\end{eqnarray*}
for $j\neq i$.

Using Eq. (\ref{eq:cjOverlap}) we can write the time evolution of the
evolution matrix $\mathbf{E}$ as follows
\begin{eqnarray}\label{EjjOverlap}
\nonumber \mathcal{E}_{jj}(T) &=& 1 -\alpha_j^2 \, \int \! d\omega J(\omega) F_{jj}(\omega,T)\\
&\approx &e^{-\alpha_j^2 \, \int \! d\omega J(\omega)
F_{jj}(\omega,T)},
\end{eqnarray}
\begin{eqnarray}\label{EjiOverlap}
\mathcal{E}_{ji}(T) &=&-\alpha_j \alpha_i \int \! d\omega
J(\omega) F_{ji}(\omega,T) \, , j\ne i.
\end{eqnarray}

Equations (\ref{EjjOverlap})-(\ref{EjiOverlap}) show that the
short-time non-exponential behavior of the survival amplitudes
$c_j(T)$, and so of the decoherence, is crucially determined by
the way in which the qubit-reservoir coupling is modified by the
form factors, $F_{jj}(\omega,T)$ and $F_{ji}(\omega,T)$
\cite{kof00}. These terms are generally functions of $\omega$
sharply peaked
 around $\omega_{j,i}$.
We will see in next section how frequent measurements, modifying
the form factors, affect the dynamics of both quantum and
classical correlations.

\section{Zeno and anti-Zeno effects on classical an quantum correlations}

In this section we study the effect of measurements on the
dynamics of correlations focusing on the off-resonant regime.
Indeed, in these conditions, the dynamics of entanglement is much
reacher than in the resonant case \cite{fran09}, and we expect
that measurements may cause both quantum Zeno and anti-Zeno
effects. On the contrary, as shown in Ref. \cite{man08}, only the
Zeno effect occurs on resonance.

\subsection{Zeno dynamics of the survival amplitude}

The analytical expressions of concurrence, classical correlations
and quantum discord, as given by Eq. (\ref{Ct}), Eq.
(\ref{classCorr}), and Eq. (\ref{QuantDisc}), respectively, all
depend on the modulus of the survival amplitudes $\vert
c_{1,2}^{(N)}(T) \vert$ in presence of $N$ measurements. These
quantities can be written as follows
\begin{equation}\label{ProCj}
\abs{c_j^{(N)}(T)} = \abs{\mathcal{E}_{jj}^{(N)}(T) \, c_{j0}+
\mathcal{E}_{ji}^{(N)}(T) \,  c_{i0}}.
\end{equation}
The first term of the sum within the modulus describes the
behavior of the excitation initially present in the state
$\ket{\psi_j}$ (and therefore, it is proportional to $c_{j0}$).
This is an \emph{excitation-trapping contribution}. The second
term, instead, is an \emph{excitation-transfer contribution}, as
it describes an excitation transfer from the state $\ket{\psi_i}$
to the state $\ket{\psi_j}$.

From Eq. (\ref{ProCj}) one immediately understands that, for
$c_{10}\neq 0$ and $c_{20}\neq 0$, the survival probability
$P_j^{(N)}(T)= \abs{c_j^{(N)}(T)}^2$ is crucially dependent on an
interference term proportional to $\Re
\Big[\mathcal{E}_{jj}^{(N)}\mathcal{E}_{ji}^{(N)}\Big]$. This, in
turns, implies that the relative phase $\phi$ between the two
components of the initial two-qubit state plays a key role.
Indeed, we will see how the correlation dynamics is strongly
sensitive to the specific initial state. In particular we will
show that initial pure states possessing the same degree of
entanglement, such as two types of initial Bell-like states
corresponding to $\phi=0$ and $\phi=\pi$, display very different
qualitative dynamics of correlations in presence of measurements.

In the bad cavity limit considered in this paper, we can further
approximate Eqs. (\ref{EvoNsum1})-(\ref{EvoNsum2}) as follows
\begin{eqnarray} \label{EvoNsum-approx1}
\mathcal{E}_{jj}^{(N)}(T) &\approx& \mathcal{E}_{jj}^N(T), \\
\label{EvoNsum-approx2}
\mathcal{E}_{ji}^{(N)}(T) &\approx& \mathcal{E}_{jj}^N(T)
\frac{\mathcal{E}_{ji}(T)}{\mathcal{E}_{jj}(T)}\sum_{k=0}^{N-1}\left(\frac{\mathcal{E}_{ii}(T)}{\mathcal{E}_{jj}(T)}
\right)^k.
\end{eqnarray}

Using Eqs. (\ref{EjjOverlap})-(\ref{EjiOverlap}), Eq.
(\ref{ProCj}), and Eqs.
(\ref{EvoNsum-approx1})-(\ref{EvoNsum-approx2}), a straightforward
calculation allows us to recast the survival amplitudes in
presence of measurements in the form

\begin{eqnarray}\label{ProCj-approx}
 \!\!\! \abs{\, c_j^{(N)}(T)} & \approx & e^{-\gamma_{jj}(T) \, t} \, \abs{ \, c_{j0} + \frac{\mathcal{E}_{ji}(T)}{T} \; \epsilon_{ji}(T,t) \, c_{i0} },
\end{eqnarray}
where
\begin{eqnarray}
  \epsilon_{ji}(T,t) &=& \frac{e^{\left[\gamma_{jj}(T)-\gamma_{ii}(T)+ i\left(\phi_{jj}(T)-\phi_{ii}(T)\right)\right]\, t}-1}{\gamma_{jj}(T)-\gamma_{ii}(T)+ i\left(\phi_{jj}(T)-\phi_{ii}(T)\right)}, \label{epsilon}
\end{eqnarray}
for $\omega_1\neq \omega_2$, and
 \begin{eqnarray}
  \epsilon_{ji}(T,t) &=& \frac{t}{\mathcal{E}_{jj}(T)},
\end{eqnarray}
for $\omega_1= \omega_2$.
The phase factor appearing in Eq. (\ref{epsilon}) is given by
\begin{eqnarray}\label{arg}
   \nonumber \phi_{jj}(T)&=&\frac{1}{T} \Im \, \int_0^\infty \! d\omega J(\omega) F_{jj}(\omega,T),
\end{eqnarray}
while the effective decay rate in presence of measurements is
\begin{eqnarray}\label{decayRate}
   \nonumber \gamma_{jj}(T)&=&\frac{1}{T} \Re \, \int_0^\infty \! d\omega J(\omega) F_{jj}(\omega,T)\\
   &=& \frac{T}{2} \int_0^\infty \! d\omega J(\omega) \, \mathrm{sinc}^2 \left[\frac{(\omega_j-\omega)T}{2}
   \right].
\end{eqnarray}
The effective Zeno decay rate is the overlap integral of the
measurements-induced atomic level broadening of width $\nu=1/T$,
described by $F_{jj}(\omega,T)$ and the environmental spectrum
$J(\omega)$. Depending on the form of the reservoir spectrum, the
effective decay can be enhanced or inhibited, for short $T$, with
respect to the dynamics in absence of measurements, giving rise to
the anti-Zeno or quantum Zeno effects, respectively  \cite{kof00}.

In particular, for the Lorentzian spectrum considered here, and in
the near-resonant regime, one can see from Eq. (\ref{decayRate})
that the effective decay rate decreases with decreasing $T$, since
$T \ll \lambda^{-1}, \delta_j^{-1}$  \cite{kof00}. In the far
off-resonant regime, on the contrary, there exist values of $T$
short enough to satisfy the conditions of validity of our
theoretical description, but at the same time such that $T \gg
\delta_j^{-1}$. In this case, one can see from Eq.
(\ref{decayRate}), that the effective decay rate increases with
respect to the value in absence of measurements, and it keeps
increasing for increasing values of $T$. This measurements-induced
enhancement of the decay is a signature of the anti-Zeno effect.
Summarizing, the Zeno decay rate is reduced in the near-resonant
regime, while in the off-resonant regime an enhancement of the
decay rate can occur since generally $T \ll \delta_j^{-1}$.

The behavior of the effective decay rate of atom $j$, therefore,
strongly depends on the position of its atomic frequency with
respect to the peak of $J(\omega)$. However, Eq.
(\ref{ProCj-approx}) tells us that the dynamics of $
\abs{c_j^{(N)}(T)}$ is influenced also by the excitation transfer
contribution $\mathcal{E}_{ji}(T)$ which in turn depends on the
position of the Bohr frequency of the other atom $i$ with respect
to $J(\omega)$. As we will see in the following, the excitation
transfer contribution is responsible for the appearance of
oscillations between Zeno and anti-Zeno behavior.

\subsection{Zeno dynamics of quantum correlations}
We now focus on the dynamics of quantum correlations for equal
couplings between the two atoms and the reservoir, i. e., when
$r_1=r_2$, and  in the two regimes $\delta_1=\delta_2$ and
$\delta_1=-\delta_2$, with $\delta_1,\delta_2 \neq 0$, for which,
as reminded above, the free dynamics of quantum correlations shows a
peculiar behavior \cite{fran09}.

The two cases present distinctive features. For
$\delta_1=\delta_2$ the free dynamics of the concurrence is
strongly dependent on the initial condition. This is due to the
existence of a subradiant state. Therefore, the time evolution of
the two orthogonal maximally entangled states, $s=0$ and
$\phi=\pi,0$, is very different. The $\phi=\pi$ initial state is
subradiant and does not evolve in time, while the $\phi=0$ state
is coupled to the cavity field and its initial entanglement, for
short initial times, monotonically decays,  as one can see from
Fig.1 (a).
\begin{figure}
\label{uffiguuno}
\includegraphics[width=0.45\textwidth]{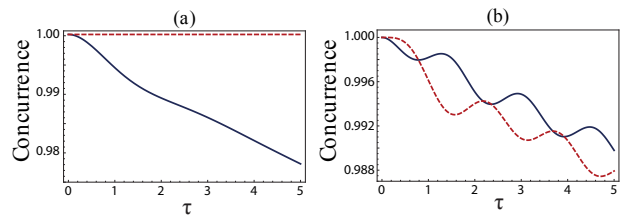}
\caption{(Color online) Short time evolution of the two-qubit
entanglement for the two initial phases $\phi=0$ (dashed lines)
and $\phi=\pi$ (solid lines), for the two regimes (a)
$\delta_1=\delta_2$ and (b) $\delta_1=-\delta_2$.}
\end{figure}
However, in the presence of measurements, this dependence on the
initial state gets substantially washed out (see below).

For $\delta_1= - \delta_2$, on the contrary, the free long time
dynamics is independent on the initial condition as discussed in
detail in Ref. \cite{fran09}. However short time oscillations in
the concurrence occur, and these oscillations again depend on the
initial state. In Fig. 1 (b) we compare the short
time entanglement dynamics of the two maximally entangled states
corresponding to $s=0$ and $\phi=\pi,0$. Note that, in both cases
the initial entanglement loss is non-monotonic, contrarily to the
case  $\delta_1=\delta_2$ shown in Fig. 1 (a). These
initial oscillations in entanglement give rise to interesting
features in the time evolution of the quantum and classical
correlations in presence of measurements. Indeed, the initial short-time differences gets amplified by the measurements,
giving rise to the qualitatively different behaviors of Figs. (\ref{figuratre}) and (\ref{figuronaquattro}),
where both the quantum and classical correlations in presence of
measurement are compared with their measurement-free counterparts. In these figures show the time evolution of both both types of correlation as a function of the measurement interval
$T$ in the bad cavity limit (we have chosen the ratio between Rabi
frequency and cavity line-width to be $R={\cal R}/\lambda=0.1$),
as the effects we want to emphasize are better displayed in such a
case (see below for the details).

\begin{figure}[h]
\begin{center}
\includegraphics[width=0.43\textwidth]{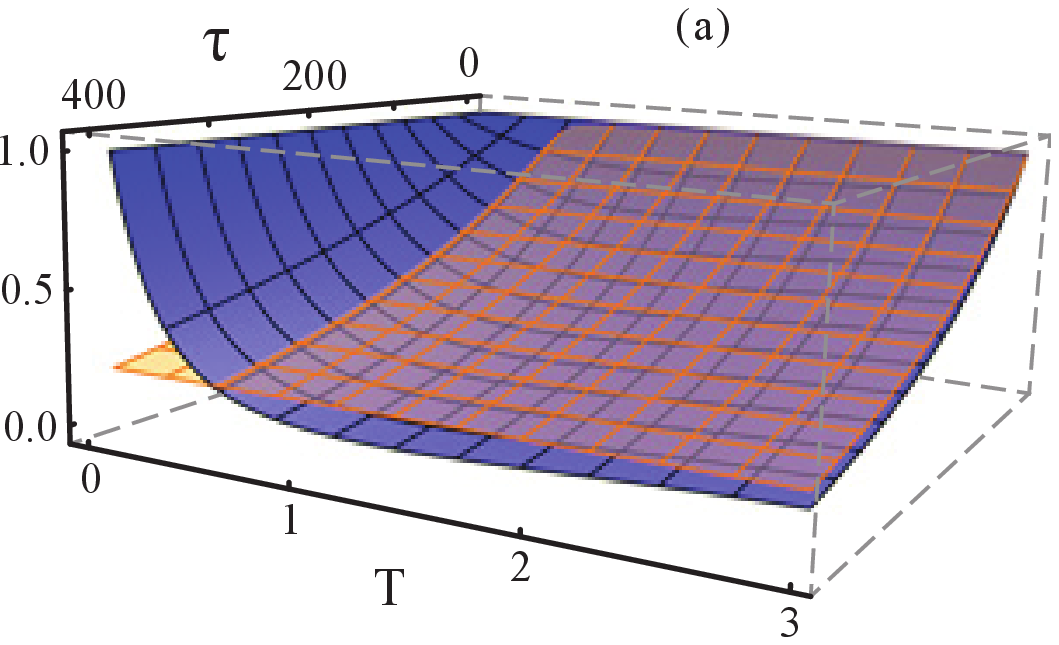}
\includegraphics[width=0.43\textwidth]{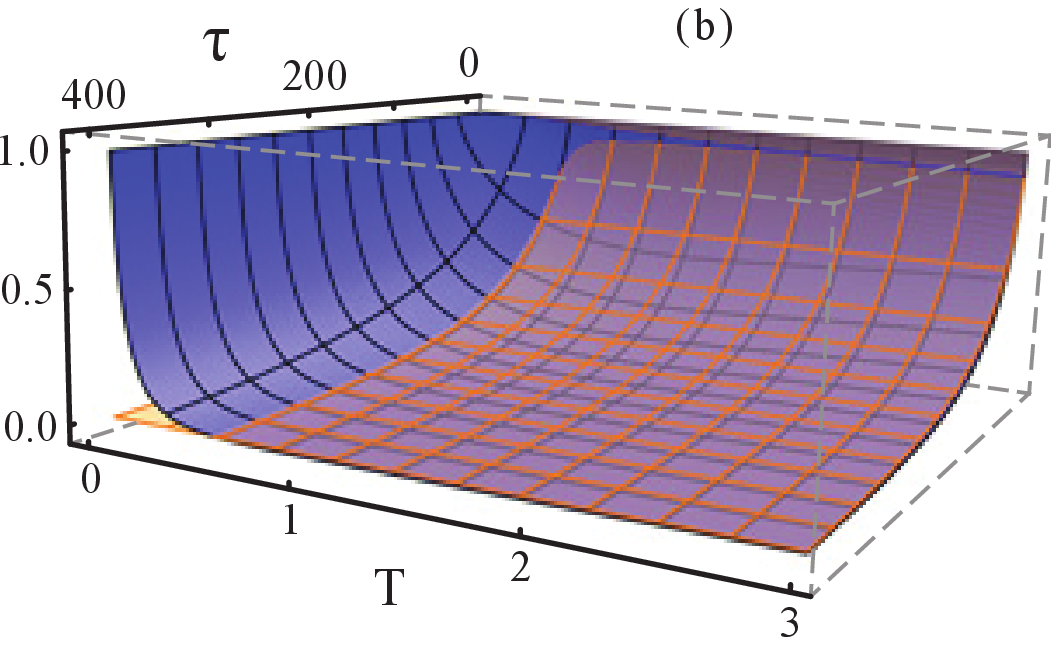}
\end{center}
\caption{(Color online) Concurrence (a), and Classical
correlations (b), in the case of equal detunings
$\delta_1=\delta_2$. Both of the functions are shown in presence
of measurements (blue surface, the upper one for small enough $T$)
and in absence of measurements (free evolution, $T$-independent,
transparent-orange surface, lying below the other one near $T=0$),
as a function of time $\tau = \lambda t$ and of the measuring
interval $T$ (also measured in units of $1/\lambda$). The plots
are performed in the bad cavity limit $\left(R=0.1\right)$, for
$r_1= 1/\sqrt{2}$, and for an initial maximally entangled state,
$s=0$, with $\phi=0$. } \label{figurdue}
\end{figure}

\begin{figure}[h]
\begin{center}
\includegraphics[width=0.43\textwidth]{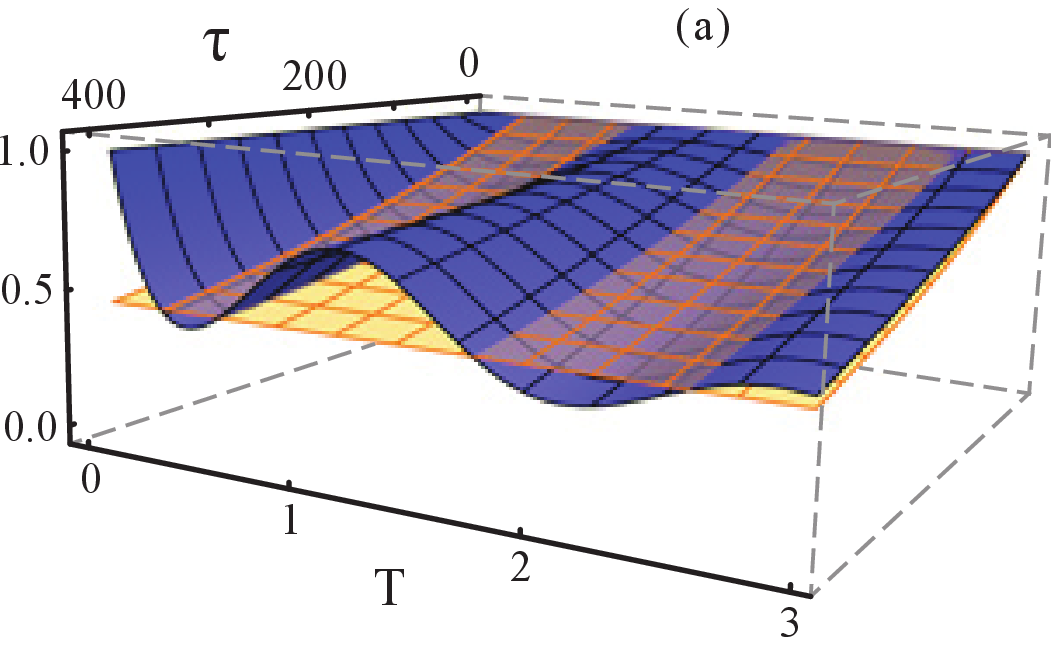}
\includegraphics[width=0.43\textwidth]{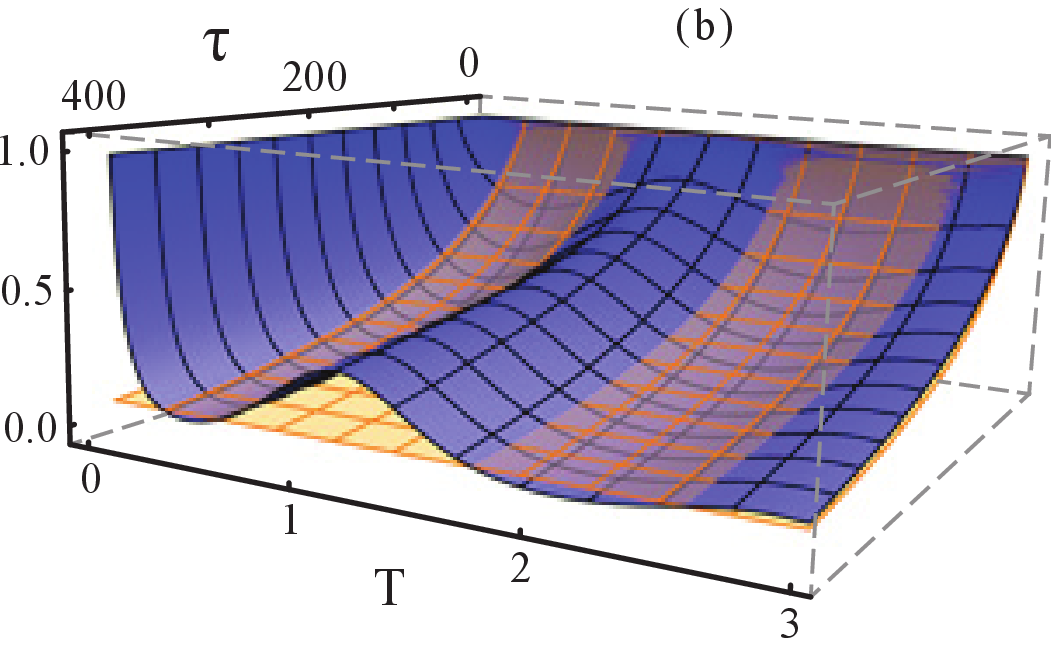}
\end{center}
\caption{(Color online) Quantum (a), and Classical correlations
(b), for symmetric detunings, $\delta_1= - \delta_2 = 2 \lambda$.
Both of them are shown in presence of measurements (blue surface,
oscillating as a function of $T$) and in absence of measurements
($T$-independent, transparent-orange surface) as a function of
time $\tau = \lambda t$ and of the measuring interval $T$ (also
measured in units of $1/\lambda$). The other parameters are $r_1=
1/\sqrt{2}$, $\left(R=0.1\right)$, $s=0$, with $\phi=0$.}
\label{figuratre}
\end{figure}

\begin{figure}[h]
\begin{center}
\includegraphics[width=0.43\textwidth]{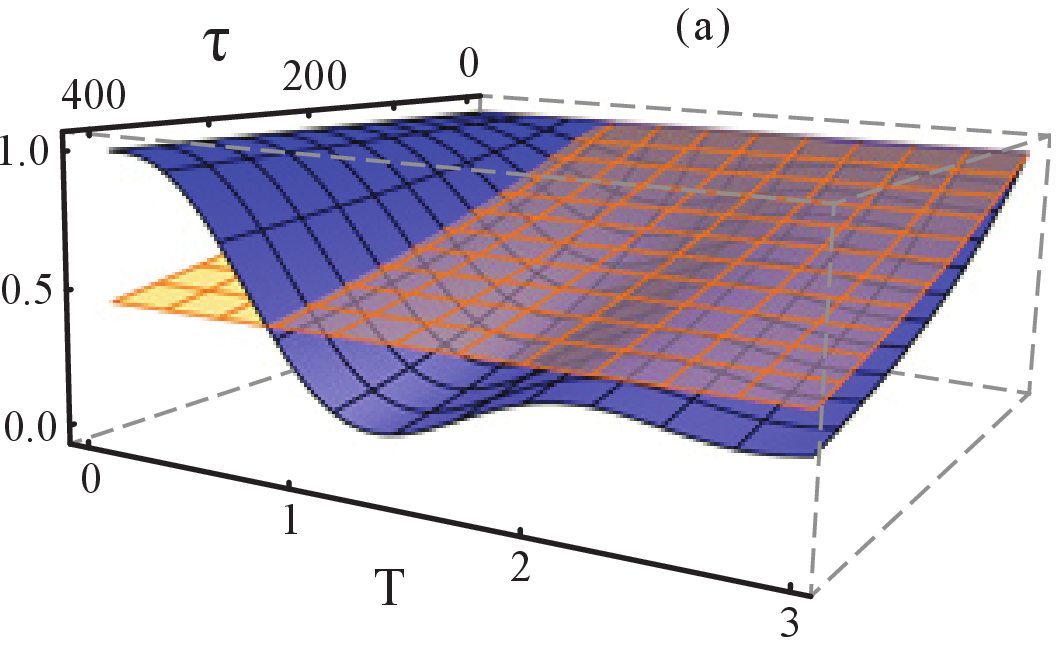}
\includegraphics[width=0.43\textwidth]{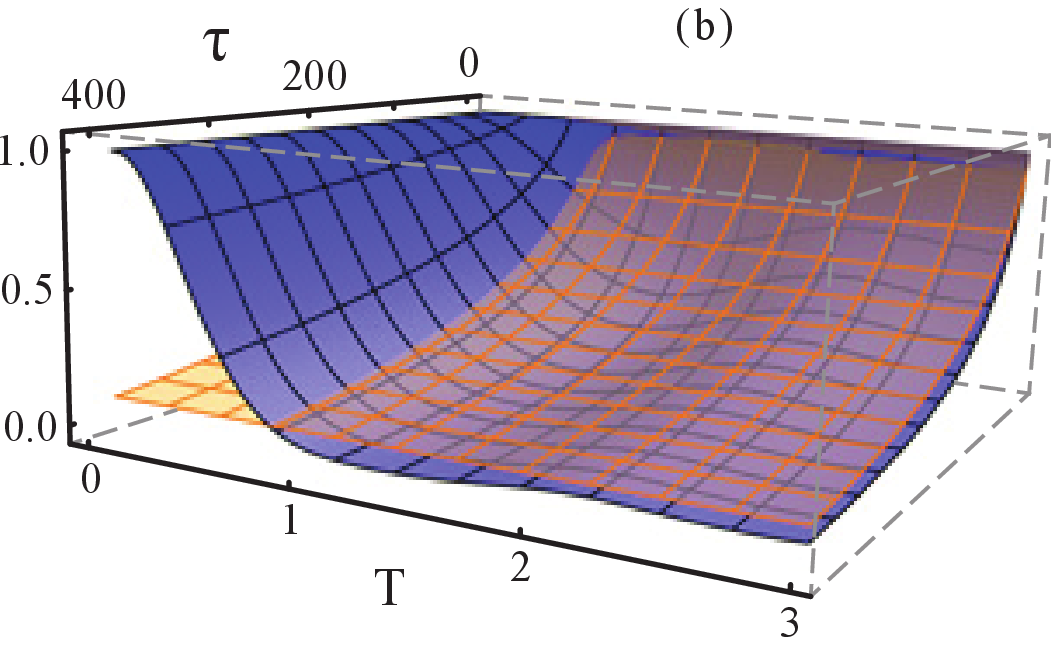}
\end{center}
\caption{(Color online) Same as Fig. \ref{figuratre}, but with a
different relative phase between the two amplitudes of the initial
states. In this case, we have chosen $\phi=\pi$.}
\label{figuronaquattro}
\end{figure}

For $\delta_1=\delta_2$, an anti-Zeno effect appears for values of
$T$ larger than a characteristic threshold value $T^*$ that
depends on the detuning and on the reservoir width, as shown in Fig
\ref{figurdue}-(a). In particular, for increasing values of the
detuning, the Zeno region becomes smaller and smaller, and the
protection against the decay occurs only for very short delay
between measurements.

On the other hand, for symmetric detunings $\delta_1=-\delta_2$,
the correlations in presence of measurements show oscillations as
a function of the measurements time interval $T$, so that quantum
Zeno and anti-Zeno effects for the entanglement alternatively
occur for increasing values of $T$, as shown both in Fig.
\ref{figuratre} and Fig. \ref{figuronaquattro}. These two figures
differ only because a different choice has been made of the
relative phase $\phi$ of the two amplitudes of the initial state.
This peculiar $\phi$-dependence is found to occur only for
symmetric detunings, $\delta_1=-\delta_2$, and is reminiscent of
the difference in the short time behaviors shown in Fig. 1.

To understand these features, we start by writing down the
concurrence in presence of measurements, $C_E^{(N)}(t)$, in the
form
\begin{eqnarray}
\nonumber C_E^{(N)}(t) &\approx &  2 \, e^{-\gamma_{11}(T)
\, t} \Big\vert  c_{10}+\frac{\mathcal{E}_{12}(T)}{T} \, \epsilon_{12}(T,t) c_{20}\Big\vert\\
& &\times e^{-\gamma_{22}(T) \, t}\Big\vert
c_{20}+\frac{\mathcal{E}_{21}(T)}{T} \, \epsilon_{21}(T,t)
c_{10}\Big\vert.
\end{eqnarray}

In the bad cavity limit, and for $\delta_1= \pm \delta_2$, this
expression can be further simplified so that the concurrence
(coinciding with the discord) is given by

\begin{eqnarray}\label{approxConcDisc}
\nonumber C_E^{(N)}(t) &\approx &  2 \left[ \abs{c_{10}}^2
  + 2 \abs{c_{10}} \abs{c_{20}}\abs{\frac{\mathcal{E}_{12}(T)}{T}}\, t \cos \theta_{12}
  \right] \\
& & \times \, e^{-2\gamma_{11}(T) \, t}, \end{eqnarray} where
\begin{eqnarray*}
  \theta_{12} &\approx & \arg \left(\mathcal{E}_{12}(T) \, c_{20}\right).
\end{eqnarray*}
The equations above explicitly show that the appearance of oscillations on the
correlations dynamics is due to the interference between
excitation-trapping and excitation-transfer contributions, which
is mainly determined by the terms $\mathcal{E}_{ji}(T)$ of Eq
(\ref{EjiOverlap}). Indeed, it is the value of $\theta_{12}$ which
is responsible for the oscillations in Eq. (\ref{approxConcDisc}),
and this angle, in turn, is crucially determined by
$\mathcal{E}_{12}(T)$ and by the relative phase $\phi$ between the
two amplitudes of the initial state.

When the two qubits have the same frequency ($\delta_1=\delta_2$),
the cosine is always positive, so that the interference term
decreases without showing oscillations. This implies that also the
dependence on the phase $\phi$ becomes almost irrelevant. On the
contrary, for qubits symmetrically detuned from the cavity mode,
the sign of the cosine changes in time, giving rise to the
observed oscillations between the Zeno and anti-Zeno regimes.
Furthermore, since $\mathcal{E}_{ji}(T)$ is divided by the
measuring interval $T$, the oscillations get amplified for
frequent measurements. These features mainly depend on the \lq positions\rq of the Bohr
frequencies of the two atoms with respect to the cavity spectrum,
which enter $\mathcal{E}_{ji}(T)$ through the off-diagonal form
factor $F_{ji}(\omega, T)$, see Eq. (\ref{EjiOverlap}).

Since $\mathcal{E}_{ji}(T)$ describes an excitation-transfer
contribution during time $0\leq t \leq T$, its behavior is
determined by the effective Rabi period of the
excitation exchange between the two atoms. In the dispersive
regime, this is of the order of the detuning and, thus, it becomes
observable in the bad cavity limit, in which one can chose a $T$
large enough to have $\delta^{-1}<T$, with a detuning larger than
the coupling strength with the cavity.

\subsection{Zeno dynamics of classical correlations}

As one can see from Figs. \ref{figurdue},
\ref{figuratre} and \ref{figuronaquattro}, classical correlations are
affected by the measurements in essentially the same way as quantum
correlations are.

In particular, ${\cal C}_c(t)$ decays in time due to the atomic
relaxation processes, but frequent enough measurement can help it
to survive. However, surprisingly enough, even the classical
correlation can display the anti-Zeno effect and, even more
surprisingly, also the oscillations between the Zeno and anti-Zeno
regimes. Furthermore, when the detunings satisfy the condition
$\delta_1=-\delta_2$, ${\cal C}_c(t)$ too shows a strong
dependence on the relative phase $\phi$ present in the initial
state, which is completely attributable to the measurement as the
measurement-free evolution of the classical correlations does not
show any dependence on $\phi$.

\section{Conclusions}
In summary, we have analyzed the dynamics of a couple of
two-level-atoms decaying in an electromagnetic resonator having a
finite quality factor and discussed, in particular, the behavior
of both quantum and classical correlations between the two atoms,
under the effect of a series of projective measurements performed,
e.g., on the cavity field. When the atoms are resonantly coupled
to the cavity, a quantum Zeno effect on the entanglement occurs,
but for off-resonant atoms, the anti-Zeno effect is obtained,
instead. Furthermore, there is a particularly distinguished regime
in which a series of oscillations occur between the Zeno and
anti-Zeno effects, as a function of the time delay between
successive measurements. We have investigated this behavior in
details, and showed a sensitivity of the coarse grained dynamics
effectively induced by the measurements to the relative phase of
the initial state. This occurs, in particular, if the atoms
interact dispersively with the electromagnetic field, having
transition frequencies which are symmetrically displaced with
respect to that of the main cavity mode. In this regime, an
analogous behavior is obtained for the classical correlations
between the atoms, which are affected by the measurements in a
qualitatively very similar way.

\begin{acknowledgments}
F.P. and S.M. acknowledge a useful discussion with prof. M. G. A.
Paris on the subject matter of this paper. This work has been
supported by the Emil Aaltonen foundation, the Finnish Cultural
foundation, and by the Turku Collegium of Science and Medicine
(S.M.).
\end{acknowledgments}

\end{document}